\pdfoutput=1
\documentclass[useAMS,usenatbib]{mn2e}
\usepackage{bm}
\usepackage{amsbsy}
\usepackage{graphicx}
\usepackage{esint}
\newcommand{\apj}{ApJ}
\newcommand{\apjl}{ApJL}
\newcommand{\jgr}{J. Geophys. Res.}
\newcommand{\aap}{A~\&~A}
\newcommand{\mnras}{MNRAS}
\newcommand{\solphys}{Sol.Phys.}
\newcommand{\physrep}{Phys. Rep.}

\title[Helicity hemispheric sign rule]
{The origin of the helicity hemispheric sign rule reversals in the mean-field solar-type dynamo}

\author[V.V.~Pipin, H.~Zhang, D.D.~Sokoloff, K.M.~Kuzanyan,  Y.~Gao]
{V.V. Pipin$^{1,2,3}$\thanks{email: pip@iszf.irk.ru},
 H. Zhang$^{1}$\thanks{email: hzhang@bao.ac.cn},
  D.D. Sokoloff$^{1,4}$\thanks{email: sokoloff.dd@gmail.com},
 K.M. Kuzanyan$^{1,5}$\thanks{email: kuzanyan@gmail.com},
 Y. Gao$^{1}$\thanks{email: gy@bao.ac.cn}\\
 $^{1}${National Astronomical Observatories, Chinese Academy of
Sciences, Beijing 100012, China}\\
 $^{2}${Institute of Solar-Terrestrial Physics, Russian Academy of
Sciences, Irkutsk, 664033, Russia}\\
 $^{3}${Stanford University, Palo Alto, CA,..... U.S.A.} \\
 $^{4}${Department of Physics, Moscow University, 119992 Moscow,
Russia}\\
 $^{5}${IZMIRAN, Russian Academy of Sciences, Troitsk, Moscow
region 142190, Russia}\\ }

\begin{document}
\label{firstpage} \maketitle
\begin{abstract}
Observations of proxies of the magnetic helicity in the Sun over the
past two solar cycles revealed reversals of the helicity hemispheric
sign rule (negative in the North and positive in the South hemispheres).
We apply the mean-field solar dynamo model to study the reversals
of the magnetic helicity sign for the dynamo operating in the bulk
of the solar convection zone. The evolution of the magnetic helicity
is governed by the conservation law. We found that the reversal of
the sign of the small-scale magnetic helicity follows the dynamo
wave propagating inside the convection zone. Therefore, the spatial
patterns of the magnetic helicity reversals reflect the processes
which contribute to generation and evolution of the large-scale magnetic
fields. At the surface the patterns of the helicity sign reversals are determined by the magnetic helicity
boundary conditions at the top of the convection zone.
We demonstrate
the impact of fluctuations in the dynamo parameters and variability in dynamo cycle amplitude
on the reversals of the magnetic helicity sign rule.
The obtained results suggest that
the magnetic helicity of the large-scale axisymmetric field can be treated as an
additional observational tracer for the solar dynamo.
\end{abstract}

\section{Introduction}

Vector magnetographic observations of the solar active regions show that
the distribution of the electric current helicity has a pronounced
anti-symmetry with respect to the solar equator
\citep{see1990SoPh,pevCa94,pev95ApJ,baoZh98,kuza00,hagsak05,zetal10}.
 This phenomenon is called the hemispheric sign rule of
current helicity. By analysis of the photospheric vector
magnetograms of active regions it has been shown that the current
helicity in the northern hemisphere is mainly negative while in the
southern hemisphere it is positive. The same hemispheric
sign rule was obtained from the synoptic magnetic field maps by
\cite{2000ApJ...528..999P} (see, also, \citealp{pevts2001ApJ}). Both
kinds of observations deal with the line-of-sight part of current
helicity, which can be identified with the total current helicity
density using the assumption of the spatial isotropy of the current
helicity distribution.

It is possible to relate the current helicity density with the magnetic
helicity density taking into account the theoretical assumption on
turbulent nature and isotropy of the magnetic fields (see, e.g., \citealp{moff:78,kle-rog99}).
Magnetic helicity is an integral of motion in MHD (\citet{wolt,moff}).
This impacts the saturation of the magnetic field generation in the
large-scale helical dynamos \citep{pouquet-al:1975a,
kleruz82,vain-kit:83,2000A&A...361L...5K,2005PhR...417....1B}.
Thus, the information about the surface distribution of the current
helicity density and about its evolution with the solar cycle may
be important for our understanding of the dynamo processes inside
the solar convection zone \citep{2003A&A...409.1097K,choud2004ApJ,2012ApJ...751...47Z}.
It is also important for understanding the processes of the magnetic
helicity transport from the convection zone to the
corona \citep{2000JGR...10510481B,warn2011,2011ApJ...734....9B}.

The observations indicate departure from the hemispheric
sign rule \citep{bao00,hagsak05}. It was found that at some periods
of the solar cycle the hemispheric sign rule reverses to the
opposite, at least at some latitudes and times \citep{zetal10}. It
was realized that the properties of these reversals may be related
with the kind of the dynamo operating in the Sun with the
distribution of the dynamo processes inside the convection zone, and
with the types of the magnetic helicity loss involved in the dynamo
(see, e.g.,
\citealp{soketal2006an,guero10,mitra11,pk11,2012ApJ...751...47Z}).
These mechanisms do not exclude the local processes which may take
part in formation of the twisted magnetic field at the subsurface
layers. Some of them were brought attention in the literature and
could be considered as alternative points of view to the problem
(see, e.g., \citealp{long1998ApJ,kps:06,pevts2007ASPC}).

The purpose of this paper is to analyze the origin of the current
helicity sign rule reversals within the framework of solar mean-field
dynamo models. In our study we examine the dynamo distributed over
the convection zone. In this model the global dynamo wave is shaped
by the subsurface shear layer \citep{pk11apjl}.
Our approach is a development of the results of the dynamo model of
\citep{pietal13apj} which alleviates catastrophic quenching by
consideration of total magnetic helicity conservation.
We compare
our results with that ones for the solar dynamo operating in overshoot
layer at the bottom of the solar convective zone (see \citealp{2012ApJ...751...47Z}).
Our study confronts the results of theoretical modeling with available
observational data from Huairou Solar Observing Station of Chinese Academy of Sciences.

\section{Basic equations}

The details of the model can be found in our previous papers (see,
e.g., \citealp{pipea2012AA,pi13r} and \citealp{pi08Gafd}, hereafter
P08). Here, we briefly outline the basic framework. We study the mean-field induction equation:
\begin{equation}
\frac{\partial\overline{\mathbf{B}}}{\partial t}=
\boldsymbol{\nabla}\times\left(\boldsymbol{\mathcal{E}}+\overline{\mathbf{U}}
\times\overline{\mathbf{B}}\right),\label{eq:dyn}
\end{equation}
where $\overline{\bm{\mathbf{U}}}$ is the mean velocity (differential
rotation); $\mathbf{\overline{B}}$ is the axisymmetric magnetic field:
\begin{equation}
\overline{\mathbf{B}}=\bm{e}_{\phi}B+\nabla\times\frac{A\bm{e}_{\phi}}{r\sin\theta},\label{eq:bm}
\end{equation}
where $\theta$ is a polar angle and $r$ is a radial distance;
$\boldsymbol{\mathcal{E}}=\overline{\bm{\mathbf{u}\times\mathbf{b}}}$
is the mean electromotive force, with $\mathbf{u}$ and $\mathbf{b}$
are being the fluctuating velocity and magnetic field, respectively.
Using the mean-field magnetohydrodynamic framework \citep{krarad80}
we write the $\boldsymbol{\mathcal{E}}$ as follows:
\begin{equation}
\mathcal{E}_{i}=\left(\alpha_{ij}+\gamma_{ij}^{(\Lambda)}\right)\overline{B}_{j}-\left(\eta_{ijk}+\eta_{ijk}^{(\delta)}\right)\nabla_{j}\overline{B}_{k},\label{eq:EMF-1}
\end{equation}
where the turbulent kinetic coefficients are: the $\alpha$ effect,
$\alpha_{ij}$; the turbulent pumping $\gamma_{ij}^{(\Lambda)}$;
the anisotropic diffusivity $\eta_{ijk}$ and the $\delta$ dynamo
effect \citep{rad69}, $\eta_{ijk}^{(\delta)}$. They depend on the
parameters of the turbulent convection, like the mean density and
turbulent diffusivity stratification, the Coriolis number $\Omega^{*}=2\tau_{c}\Omega_{0}$,
where $\tau_{c}$ is the typical convective turnover time, and $\Omega_{0}$
is the global angular velocity.

The $\alpha$ effect includes the hydrodynamic and magnetic helicity
contributions,
\begin{eqnarray}
\alpha_{ij} & = & C_{\alpha}\sin^{2}\theta\alpha_{ij}^{(H)}+\alpha_{ij}^{(M)},\label{alp2d}
\end{eqnarray}
where the hydrodynamic part of the $\alpha$-effect is defined by
$\alpha_{ij}^{(H)}$. The expressions for the turbulent kinetic coefficients
$\alpha_{ij}^{(H)}$ , $\gamma_{ij}^{(\Lambda)}$, $\eta_{ijk}$ and
$\eta_{ijk}^{(\delta)}$ are given in Appendix. The contribution of
small-scale magnetic helicity $\overline{\chi}=\overline{\bm{\mathbf{a}\cdot}\mathbf{b}}$
($\mathbf{a}$ is the fluctuating magnetic vector-potential) to the
$\alpha$-effect is defined as follows (see, P08):
\begin{equation}
\alpha_{ij}^{(M)}=2f_{2}^{(a)}\delta_{ij}\frac{\overline{\chi}\tau_{c}}{\mu_{0}\overline{\rho}\ell^{2}}-2f_{1}^{(a)}e_{i}e_{j}\frac{\overline{\chi}\tau_{c}}{\mu_{0}\overline{\rho}\ell^{2}}.\label{alpM}
\end{equation}
The principal nonlinear feedback of the large-scale magnetic field
to the $\alpha$-effect is due to a dynamical quenching because of
the generation of the magnetic helicity by the dynamo \citep{%
pouquet-al:1975a,kleruz82,2005PhR...417....1B}.
The relation of magnetic helicity on the large- and small scales \citep{2012ApJ...748...51H,pi13r}
is governed by the equation:

\begin{equation}
\frac{\partial\overline{\chi}}{\partial t}=-\frac{\partial\left(\overline{\mathbf{A}}\cdot\overline{\mathbf{B}}\right)}{\partial t}-\frac{\overline{\chi}}{R_{m}\tau_{c}}-\eta\overline{\mathbf{B}}\cdot\mathbf{\overline{J}}-\boldsymbol{\nabla\cdot}\boldsymbol{\boldsymbol{\mathcal{F}}}^{\chi},\label{eq:helcon}
\end{equation}
where ${\boldsymbol{\mathcal{F}}}^{\chi}=-\eta_{\chi}\boldsymbol{\nabla}\left(\overline{\chi}+\overline{\mathbf{A}}\cdot\overline{\mathbf{B}}\right)$
is the diffusive flux of the total magnetic helicity \citep{mitra10},
and $\eta_{\chi}$ is the turbulent diffusion coefficient for the
magnetic helicity. In the paper we use $R_{m}=10^{6}$ and $\eta_{\chi}=0.1\eta_{T}$,
where $\eta_{T}$ is the turbulent diffusivity profile (see, Appendix).
For the axisymmetric magnetic fields the large-scale magnetic vector-potential
is
\begin{equation}
\overline{\mathbf{A}}=\mathbf{e}_{\phi}T+\mathbf{r}P=\frac{\mathbf{e}_{\phi}}{r\sin\theta}A+r\mathbf{e}_{r}P.\label{eq:potent}
\end{equation}
The toroidal part of the vector potential is governed by the dynamo
equations. The poloidal part of the vector potential can be restored
from equation $\boldsymbol{\nabla}\times\left(\mathbf{r}P\right)=\mathbf{e}_{\phi}B$.
We matched the potential field outside and the perfect conductivity
at the bottom boundary with the standard boundary conditions. For
the magnetic helicity we employ $\bar{\chi}=0$ at the bottom of the
convection zone. The paper elaborates two kind of the surface boundary
conditions for the magnetic helicity:
\begin{eqnarray}
\left.\eta_{\chi}\nabla_{r}\left(\bar{\chi}+\overline{\mathbf{A}}\cdot\overline{\mathbf{B}}\right)\right|_{r=r_{e}} & = & 0,\label{eq:b1}\\
\left.\eta_{\chi}\nabla_{r}\frac{\bar{\chi}}{(\bar{\rho}\ell^{2})}\right|_{r=r_{e}} & = & 0.\label{eq:b2}
\end{eqnarray}
We call the model that satisfies the boundary conditions Eq.~(\ref{eq:b1})
as the model B1, and similar, the model B2 is referred to the Eq.~(\ref{eq:b2}).
We set the seed magnetic field of the preferred dipole parity
and with small admixture of the quadrupole one to check the parity
preference when the solution reaches the steady state.

The construction of the radial profiles for the turbulent coefficients,
which are involved in the mean electromotive force, remains rather
arbitrary for various kinds of the dynamo models. In our models
we use the solar convection zone model computed by \citet{stix:02}.
In the paper we use the same profiles for the turbulent coefficients
as in our previous papers (see, \citealp{pi13r}, and Appendix therein).

\section{Results}

\begin{figure*}
\includegraphics[width=0.99\textwidth]{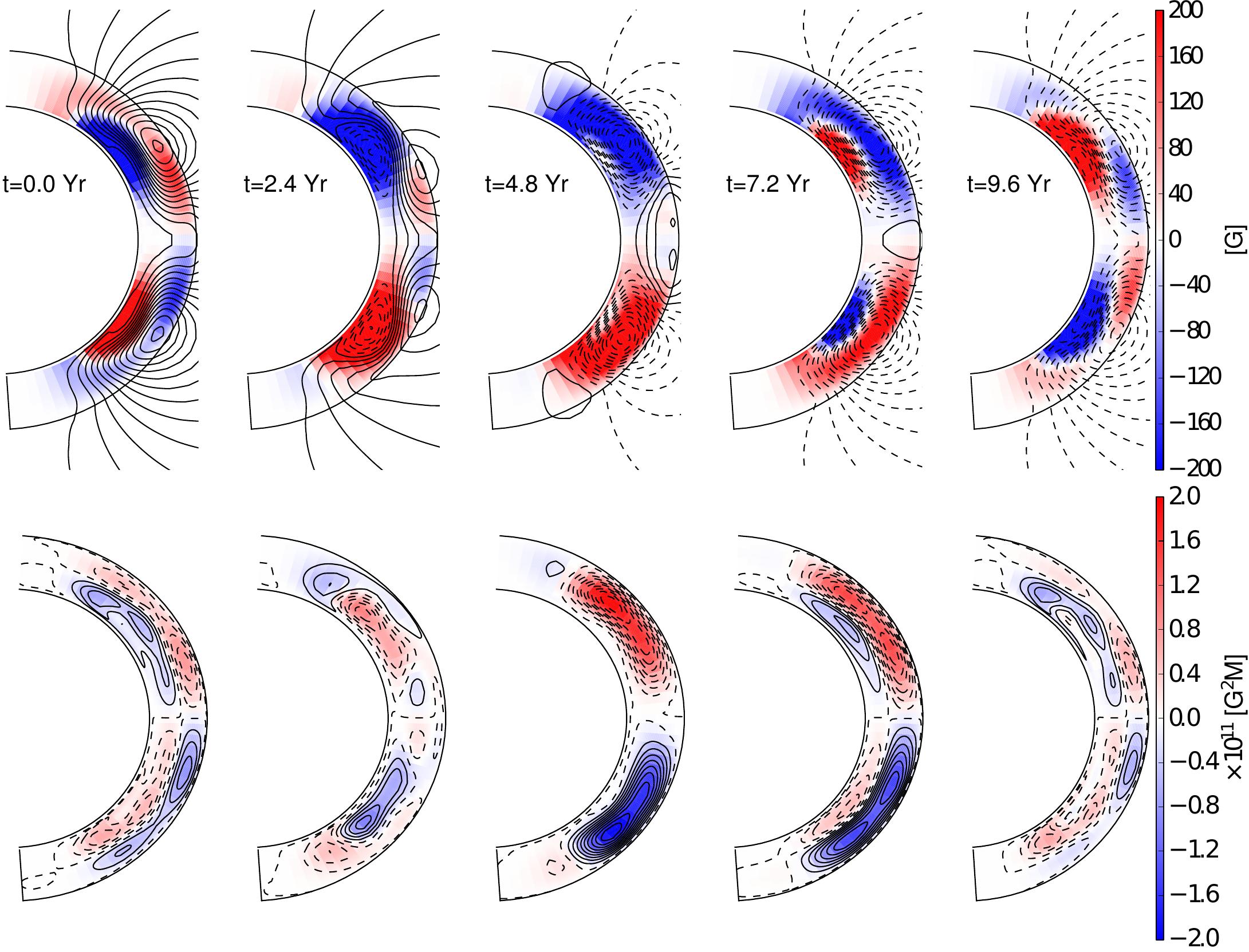}
\caption{Snapshots of the magnetic field and magnetic helicity
    evolution inside the convection zone: Top panel shows the field lines of the
poloidal component of the mean magnetic field and the toroidal
magnetic field (varies $\pm1$kG) is shown by color; Bottom
panel shows the small-scale magnetic helicity density (contours) and the
large-scale magnetic helicity density (color). Both quantities vary
with the same magnitude.
\label{fig:Snapshots}}
\end{figure*}

\begin{figure*}
\includegraphics[width=0.99\textwidth]{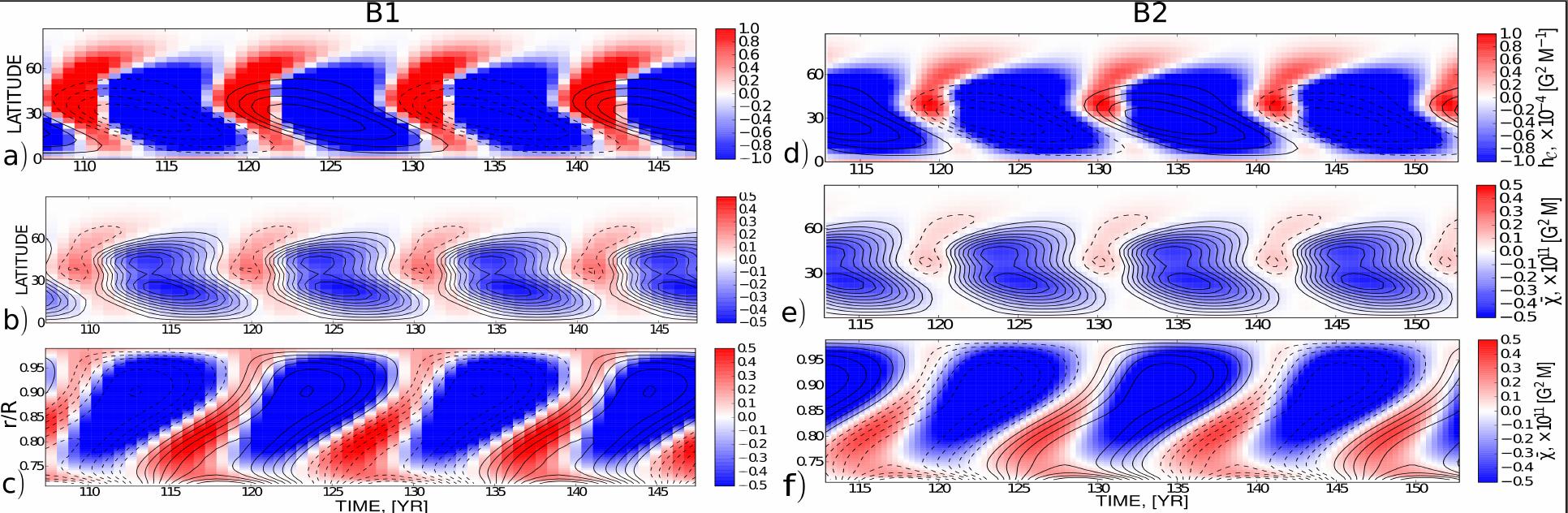}
\caption{The left column shows the results for the model B1.
Panel (a) shows the time-latitude diagram for the current helicity
(background image). Panel (b) shows the toroidal magnetic field
variations at $r=0.95R$. panel (c) shows variations of the
small-scale magnetic helicity and the toroidal magnetic field inside
the convection zone at the latitude $30^{\circ}$. Panels (d, e, f)
show the same results for the model B2. \label{fig:bfl}}
\end{figure*}

Fig.~\ref{fig:Snapshots} shows the snapshots of the magnetic field
and magnetic helicity evolution in the North hemisphere for the model
B1 which uses Eq.~(\ref{eq:b1}). The qualitatively similar results
can be obtained for the model B2. Here we see, that the spatial patterns
of the small-scale magnetic field follow the evolution of the large-scale
magnetic helicity and the latter propagates with the toroidal part
of the dynamo wave from the bottom of the convection zone to the surface.
The dynamo wave has the equatorial and the polar branches. Near the
surface the equatorial branch dominates. The hemispheric helicity
rule suggest that the small-scale helicity is negative at the North
and positive at the South hemisphere.
Fig.~\ref{fig:Snapshots}
shows that in the upper part of the convection zone the helicity rule
is valid in the most phases of the cycle. In the upper part of the
convection zone the reversal sign of the small-scale magnetic helicity
regions appears at the high latitudes when the dynamo wave of the
toroidal magnetic field comes to the subsurface shear layer. At the
equatorial latitudes the reversal sign of $\overline{\chi}$ occurs at
the decaying phases of the dynamo wave cycle. One can see that
the signs of the large and small-scale helicities are spatially related.
It is shown in Fig.~\ref{fig:bfl} that shows the
time-latitude and the time-radius variations of the magnetic field
and magnetic helicity near the surface. The Figure also demonstrates
the effect of the boundary condition change for the magnetic helicity.
For the boundary condition Eq.~(\ref{eq:b1}) the regions
with reversed sign of the small-scale magnetic helicity penetrate
into the surface while the condition Eq.~(\ref{eq:b2}) quenches this
penetration. We find that the patterns of the reversed sign of the
magnetic helicity are located at the edges of the butterfly wings
of the time-latitude diagrams for the large-scale toroidal magnetic
field.
The novel feature which is demonstrated by the Fig.~\ref{fig:bfl}
is the time-latitude diagram for the large-scale magnetic helicity
which is attributed to the axisymmetric magnetic field. It is seen
that within the current model its distribution is closely connected
with the distribution of the small-scale helicity. It is believed
that the current helicity of the surface magnetic field is the observational
proxy for the magnetic helicity $\overline{\chi}$. We note that our model
uses the full information about the large-scale magnetic helicity.

\begin{figure}
\includegraphics[width=0.48\textwidth]{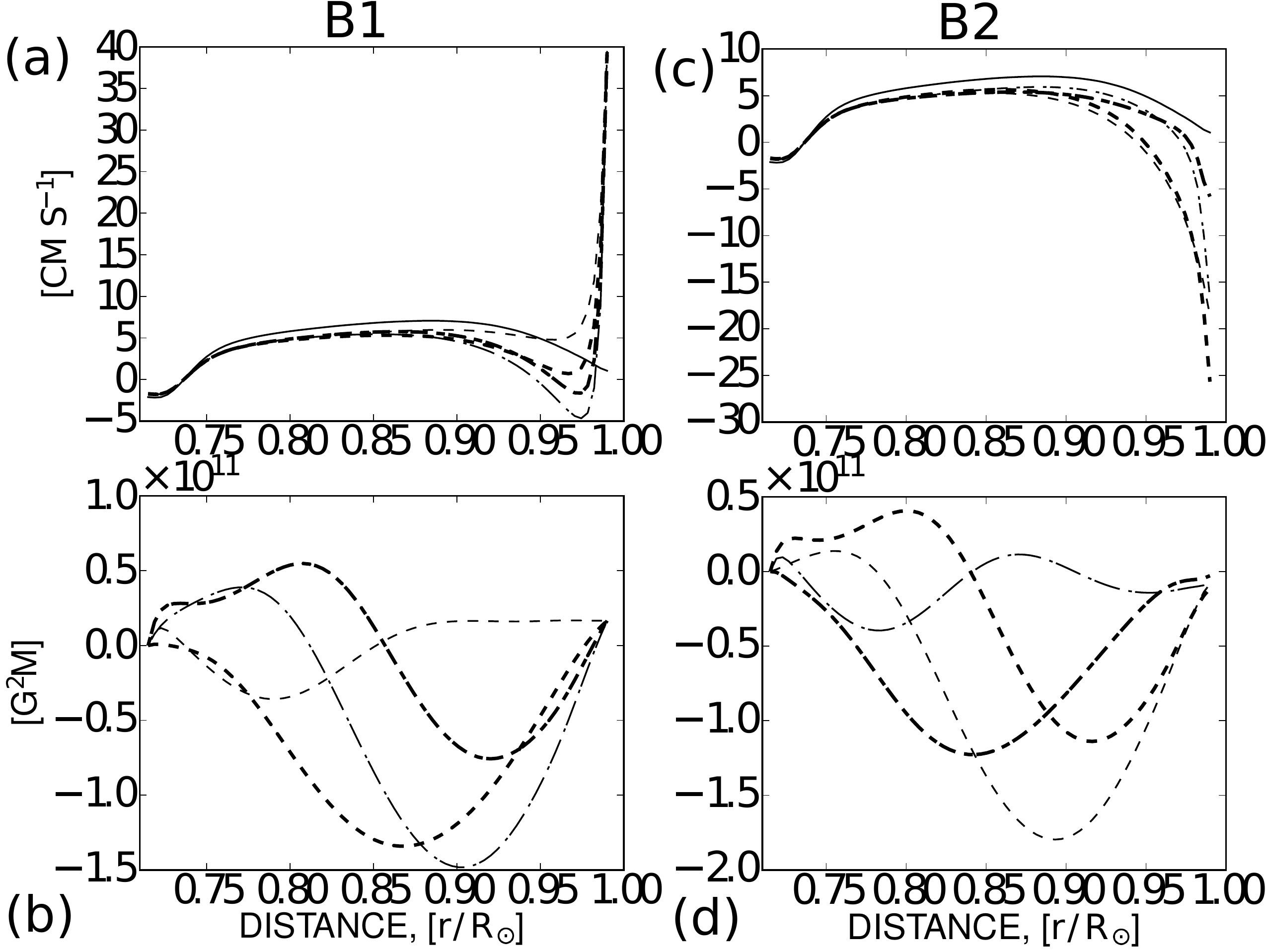}

\caption{Panels (a,b) show variations of the $\alpha$-effect and the
small-scale magnetic helicity at the latitude $45^{\circ}$ for the
model B1 (the Eq. (\ref{eq:b1})), and the panels (c,d) show the same
for the model B2 (the Eq. (\ref{eq:b2})).\label{fig:alppr}}
\end{figure}

The difference in penetration of the magnetic helicity to the surface
results in difference in the distribution of the effective $\alpha$
effect near the surface. This issue is recently discussed by \citet{kap2012},
\citet{pietal13apj}
and \citet{pi13r}. Fig.~\ref{fig:alppr} shows the snapshots of
the $\alpha_{\phi\phi}$ and the small-scale magnetic helicity profiles
for the different phase of the cycle at the latitude $45^{\circ}$.
We find that for the model B1 the $\alpha$-effect can be negative
at the certain phases of the cycle and it has the sharp positive profile
near the surface. The model B2 has the negative $\alpha$-effect for
$r>0.92R$ with the sharp negative profile near the surface. The abrupt
growth of the $\alpha$ effect amplitude near the surface is because
of the factor $\left(\overline{\rho}\ell^{2}\right)^{-1}$ in the
definition, see Eq.~(\ref{alpM}). It remains the matter of the direct
numerical simulations to justify the correct choice of the boundary
condition for the magnetic helicity. The model B2 has the zero boundary
condition for the derivative of the small-scale current helicity at
the top, see Eq.~\ref{eq:b2}. It is found that for the condition
$\left.\nabla_{r}\overline{\chi}\right|_{{\displaystyle r=r_{e}}}=0$
the negative part of the alpha-effect near the surface is stronger
than one in model B2.

\subsection{Impact of dynamo fluctuations on the helicity patterns}

\begin{figure*}
\includegraphics[width=0.99\textwidth]{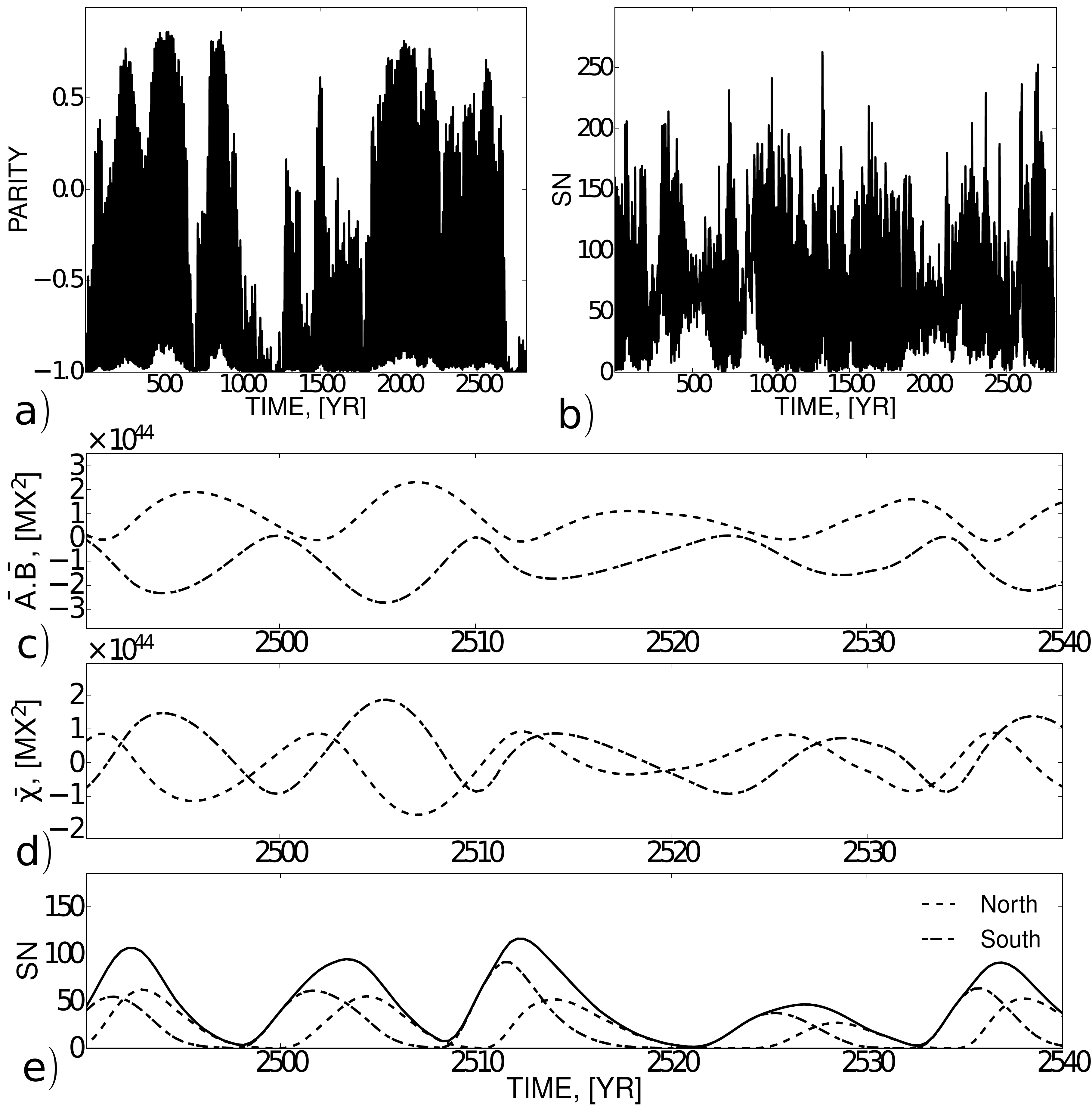}

\caption{\label{fig:fluc2}The integral characteristics of the
magnetic activity near the surface, $r=0.95R$ : a) the parity index
determining the symmetry of the toroidal magnetic field about
equator, -1 corresponds to the dipolar symmetry and 1 to the
quadrupolar; b) the simulated sunspot number {(SN)}; c) the
latitudinal integral of the large-scale magnetic helicity, dashed
line the North hemisphere, the dash-dotted line the South one; d)
the same as (c) for the small-scale magnetic helicity; e)
the same as (b), (the SN),  the solid line is the total SN, the dashed
line is the SN for the North hemisphere and the dash-dotted line is the SN for
the South hemispere.}
\end{figure*}

The sign reversals of the helicity rule can be due to random fluctuations
in the dynamo parameters and due to some random processes which generate
the magnetic helicity independent of the large-scale dynamo. In this
subsection we examine the effect of fluctuations in the dynamo parameters
on the magnetic helicity distribution variations. We exploit here
a scenario \citep{moss-sok08,uetal09,pipea2012AA} with fluctuations
of $\alpha$-effect as a possible source of the solar activity cycle
parameters from one cycle to another. We introduce random non-symmetric
about equator variations of the $\alpha$-effect,
$C_{\alpha}=C_{\alpha}\left(1+0.2\left(\xi_{N}\Theta\left(\mu\right)
+\xi_{S}\Theta\left(-\mu\right)\right)\right)$,
where $\mu=\cos\theta$, $\Theta$ is the Heaviside function and
$\left|\xi_{S,N}\right|<2\sigma$$\left(\xi_{S,N}\right)$
is the random Gaussian noise with the randomly floating phase and
with the mean memory time equals to the dynamo cycle length. In this subsection
the model B1 is discussed as it shows the stronger reversals of the
helicity rule than the model B2.

We found that the reversals of the helicity rule are stronger during
the periods of the grand minimum which are also related to the periods
of the strong hemispheric asymmetry in the magnetic activity.
Fig.~\ref{fig:fluc2} shows variations of the integral parameters of the
model for the near surface magnetic field. In our results we show
the parity index, that determines the symmetry of the toroidal magnetic
field about equator, with the value $-1$ corresponds to the dipolar symmetry of
the near surface toroidal magnetic fields
and the value $1$ corresponds to  the quadrupolar symmetry. The sunspot number was simulated in
following to \citet{pipea2012AA}. We also show the integral magnetic
helicity for each hemisphere. The magnitude of the helicity variations
is in agreement with the observational constraints obtained by \citet{2000JGR...10510481B}.
Variations of the magnetic helicity go in anti-phase at the large
and small scales, because it is prescribed by Eq.(\ref{eq:helcon}).
Nevertheless, for each hemisphere, there is a difference between the
evolution of $\mathbf{\overline{A}\cdot}\overline{\mathbf{B}}$ and
$\overline{\chi}=\overline{\mathbf{a\cdot}\mathbf{b}}$. The small-scale
helicity, $\overline{\chi}$ does change the sign in a course of the
solar cycle and the large-scale helicity $\mathbf{\overline{A}\cdot}\overline{\mathbf{B}}$
almost does not. This is similar to results shown in Fig.~\ref{fig:bfl},
where we see that reversals of the helicity rule is much stronger
for the small-scale helicity than for the large-scale one. Another
interesting results is that the maxima of the integral large-scale
magnetic helicity are approximately corresponded to the  maxima of
the decay rate in the simulated sunspot activity (cf, Figs.~\ref{fig:fluc2}{c,d}
and Fig.~\ref{fig:fluc2}e). This is due of the oscillatory character
of the dynamo and delay between the activity of the major components
of the large-scale magnetic fields which are related to the toroidal
magnetic field and the large-scale toroidal vector-potential determining the poloidal magnetic field.

\begin{figure*}
\begin{centering}
\includegraphics[width=0.8\textwidth]{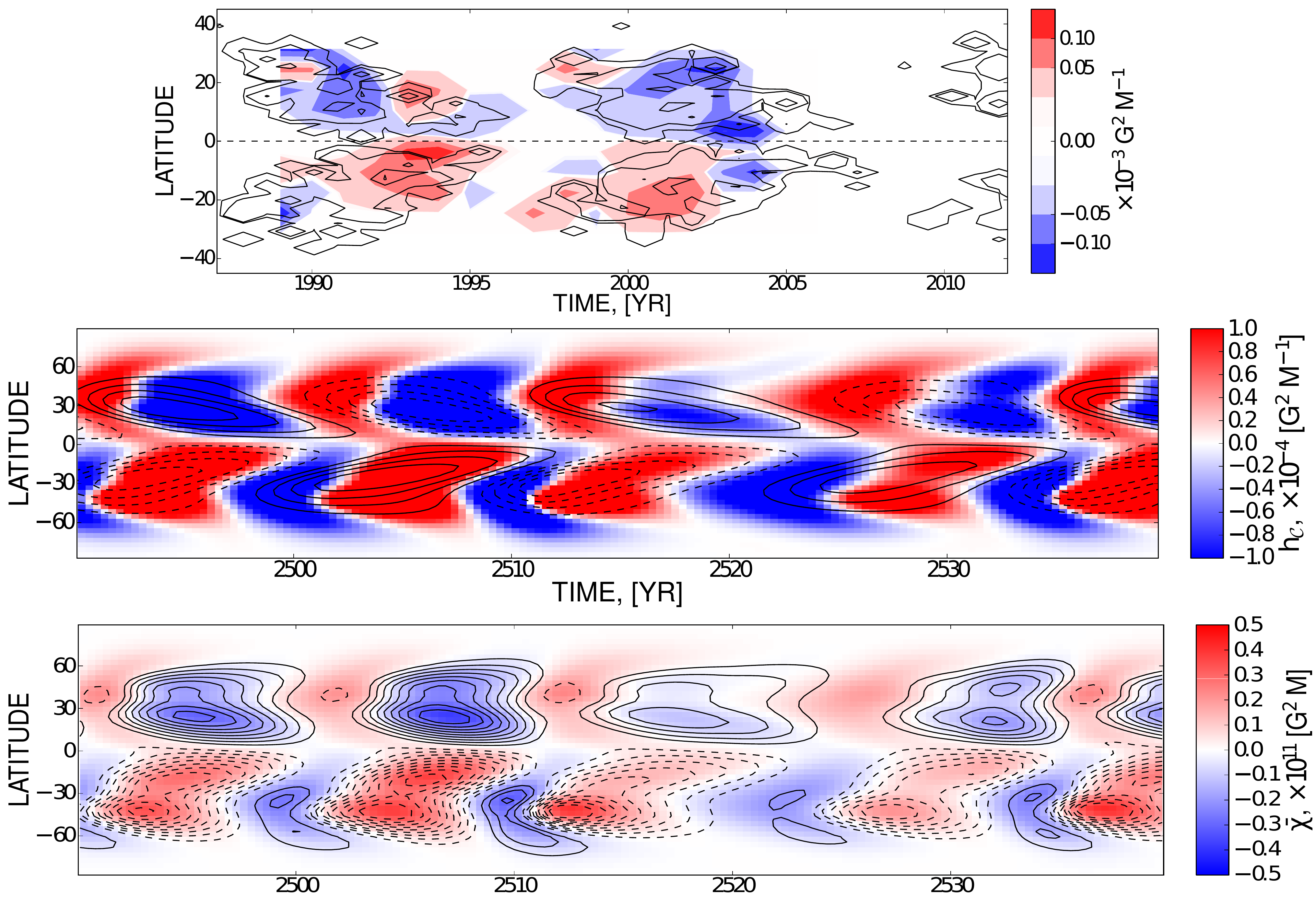}
\par\end{centering}

\caption{\label{fig:observ} The time-latitude diagram for the
current helicity as inferred from the solar active region
observations at Huairou {Solar Observing Station.} Sunspot density which traces
toroidal magnetic field is shown by colors while contour lines show
current helicity, the vertical color bar on the right side scales
the magnitude.\label{fig:fluc1}}
\end{figure*}

Finally, Fig.~\ref{fig:fluc1} shows comparison of the results for the
simulated time-latitude diagrams for the toroidal magnetic field and
the magnetic helicity with results of the current helicity observations
reported by \citet{zetal10}.
{ We used systematic series of vector magnetographic observations of solar
 active regions by 35~cm filter type SMTF telescope at Huairou Solar
 Observing Station of Chinese Academy of Sciences. The data set comprises
 6205 individual magnetograms of active regions more or less homogeneously
  covering the 18 year period of 1988-2005, which is almost two sunspot cycles.}
  The data have been grouped and averaged into statistically significant
  sub-samples in time-latitudinal bins (2 years in time and 7 degrees in
  helio-latitude), see \citet{zetal10} for details.
We have subsequently smoothed the data by using standard IDL linear
interlopation for retaining only global features of the time-latitudinal
distribution of helicity.

The results of the dynamo model are shown for the period of the grand minimum.
It is the same period as discussed for the Fig.~\ref{fig:fluc2}(c,d,e) above.
The simulated butterfly diagrams are in visible qualitative agreement with
the observations.
One can see that the model keeps the basic anti-symmetry of helicity (negative
in the North and positive in the South, i.e. the so-called hemispheric sign rule),
however with evolution it shows
various deviations from perfect periodicity (e.g., longer cycles, suppression of
activity, asymmetry in the phases of growth and decay, asymmetry in the shape of
wings on butterfly diagrams etc). It looks plausible that long and weak cycles are
associated with larger areas of helicity of the sign opposite to the hemispheric sign rule.

\section{Discussion and conclusions}

The available bulk of the current helicity data covers two activity
cycles and the transition to the following activity cycle which has been
quite unusual. The observed helicity butterfly diagrams demonstrate
that the size of the areas with the opposite helicity signs in the
later cycle differs substantially from the first one. We studied the
origins of the reversals of the magnetic helicity sign in the mean-field
solar dynamo. The evolution of the magnetic helicity in the model
subject to the global constraint of the magnetic helicity conservation
law. The nonlinear feedback of the large-scale magnetic field to the
$\alpha$-effect is described by dynamical quenching due to the
constraint of magnetic helicity conservation. The magnetic helicity,
$\overline{\chi}$, is subjected to the conservation law.

In the model, the sign reversals of the small-scale magnetic helicity
are always related with the sign reversals of the large-scale magnetic
helicity. This is due to the magnetic helicity conservation constraint.
The result develops the simple model by \citet{2009ARep...53..160X}.
The idea was recently elaborated by \citet{2012ApJ...751...47Z} for the toroidal part of the current helicity.
Our model employs the total large-scale magnetic helicity and not
only its toroidal part. Taking into account Eqs.(\ref{eq:bm},\ref{eq:potent})
we get the large-scale magnetic helicity formula for the spherical
coordinates:
\begin{eqnarray}
\overline{\mathbf{A}}\cdot\overline{\mathbf{B}} & = & \frac{A\, B}{r\sin\theta}+\frac{P}{r\sin\theta}\frac{\partial A}{\partial\theta},\label{LSH}
\end{eqnarray}
where, ${\displaystyle B=\overline{B}_{\phi}=-\frac{\partial P}{\partial\theta}}$
and ${\displaystyle \overline{B}_{r}=\frac{1}{r^{2}\sin\theta}\frac{\partial A}{\partial\theta}}$.
We have to notice that the magnetic helicity of the large-scale axisymmetric
field can be restored from observational tracers of $\overline{B}_{\phi}$
and $\overline{B}_{r}$ either from the vector magnetograms \citet{see1990SoPh} or from
the line-of-sight magnetic observations, e.g., using method by \citet{pevts2001ApJ}.
The toroidal part of the potential can be restored from the surface
distribution of the $B_{r}$ as $A\left(\theta\right)=\int_{0}^{\theta}r^{2}\sin\theta\overline{B}_{r}d\theta$
which is equivalent to the flux going outside of the Sun, and, similar,
we can restore poloidal part of vector-potential using
$P=-\int_{0}^{\theta}\overline{B}_{\phi}d\theta$. Note, that the total helisity
remains zero because of the equatorial symmetry of the axisymetric magnetic field. Hence, the
using of the Coulomb gauge is justified and the determination of the
latitudinal distribution of the  magnetic helicity of the
axisymmetric large-scale magnetic fiel is unique.

Therefore, the observations can give an information about the magnetic
helicity of the large-scale magnetic fields of the Sun.
Our results indicate (see, Figs.~\ref{fig:fluc2}(c,d,e)
that the reversal of magnetic helicity and lower values of integral helicity may proceed the lower amplitude of cyclic dynamo activity.

We found that in the models the wave of the reversed magnetic helicity
sign propagates from the bottom of the convection zone. Similar property
was recently found by \citet{warn2011} in direct numerical
simulations. The models B1 and B2 show the possibility as for the
strongly positive as well for the negative dynamical $\alpha$-effect near
the top of the convection zone. The model employs the subsurface rotational
shear having the negative radial gradient of the angular velocity.
Therefore, in case B1 the dynamo wave penetrates closer to equator
than the model B2 because of the Parker-Yoshimura rule \citep{1955ApJ...121..491P,yosh1975}.
The numerical simulations demonstrate a similar effect \citep{kap2012}.

Generally, the sign reversals of the magnetic helicity are stronger
in the model B1 than in the model B2. These two models have different
boundary conditions for the magnetic helicity at the top. In the model
B1 the diffusive flux of the large-scale helicity from the surface is balanced
by a counterpart from the small-scale helicity. Therefore, integrating
the Eq.~(\ref{eq:b1}) from some level $r=r_{0}$ the top $r=r_{e}$
we have $\overline{\chi}_{e}=\overline{\chi}_{0}+\overline{\mathbf{A}}_{0}\cdot\overline{\mathbf{B}}_{0}$,
where we use the magnetic boundary conditions as well. Thus, in the
model B1, the boundary conditions support the penetration of the local
helicity
$\overline{\chi}_{0}+\overline{\mathbf{A}}_{0}\cdot\overline{\mathbf{B}}_{0}$
(governed by Eq.~\ref{eq:helcon}) from depth to the surface. For
the boundary condition Eq.~(\ref{eq:b2}) we have $\nabla_{r}h_{\mathcal{C}}=0$
at the top. This is the same as $\nabla_{r}\overline{\chi}=\overline{\chi}\Lambda(\bar{\rho}\ell^{2})$,
where $\Lambda(\bar{\rho}\ell^{2})=\nabla_{r}\log(\bar{\rho}\ell^{2})$.
Thus, at the near surface level, in the model B2 the small-scale magnetic
helicity is determined by the profile of $\Lambda(\bar{\rho}\ell^{2})$
and not by Eq.(\ref{eq:helcon}). The further study requires clarification of
the issue if the boundary conditions impact the sign reversals
of the magnetic helicity. We can make conjecture that the change in the
boundary conditions results in the larger or smaller sign reversals
of the magnetic helicity at the surface.

The main results of the paper can be summarized as follows. The current
model suggests that the reversal of the sign of the small-scale magnetic
helicity follows the dynamo wave propagating inside the convection
zone. This was also suggested  by the numerical simulations
\citep{warn2011}.
Therefore, the spatial patterns of the magnetic helicity reversals
reflect the processes which contribute to generation and evolution
of the large-scale magnetic fields. At the surface the patterns of
the helicity rule reversals are determined by the magnetic helicity
boundary conditions at the top of the convection zone. The model suggests
that the magnetic helicity of the large-scale axisymmetric field can
be used as an additional observational tracer for the solar dynamo.

\section*{Acknowledgments}

V.P., D.S. and K.K. would like to acknowledge support from Visiting
Professorship Programme of Chinese Academy or Sciences 2009J2-12 and
thank NAOC of CAS for hospitality, as well as acknowledge support
from the collaborative NNSF-RFBR grant 13-02-91158, and RFBR grants
12-02-00170-a, 13-02-01183-a, the support of the Integration Project
of SB RAS N 34, and support of the state contracts 02.740.11.0576,
16.518.11.7065 of the Ministry of Education and Science of Russian
Federation. H.Z. would like to acknowledge support from National
Natural Science Foundation of China grants: 41174153 and 10921303.
 Y.G. would like to acknowledge support from National Natural Science
Foundation of China grants: 11103037.


\section*{Appendix}
Here we describe some details of the dynamo model that can be also
fond in \citep{pipea2012AA,pi13r}) and \citep{pi08Gafd}(hereafter
P08).
The hydrodynamic part of the tensor $\alpha_{ij}$ is represented by
$\alpha^{(H)}_{ij}$ (P08):
\begin{eqnarray}
\alpha_{ij}^{(H)} & = & \delta_{ij}\left\{ 3\eta_{T}\left(f_{10}^{(a)}\left(\bm{e}\cdot\boldsymbol{\Lambda}^{(\rho)}\right)+f_{11}^{(a)}\left(\bm{e}\cdot\boldsymbol{\Lambda}^{(u)}\right)\right)\right\}\label{eq:alpH}\\
 & + & e_{i}e_{j}\left\{ 3\eta_{T}\left(f_{5}^{(a)}\left(\bm{e}\cdot\boldsymbol{\Lambda}^{(\rho)}\right)+f_{4}^{(a)}\left(\bm{e}\cdot\boldsymbol{\Lambda}^{(u)}\right)\right)\right\} \nonumber \\
 &+ & 3\eta_{T}\left\{
   \left(e_{i}\Lambda_{j}^{(\rho)}+e_{j}\Lambda_{i}^{(\rho)}\right)f_{6}^{(a)}\right. \nonumber\\
 &+ & \left. \left(e_{i}\Lambda_{j}^{(u)}+e_{j}\Lambda_{i}^{(u)}\right)f_{8}^{(a)}\right\} ,\nonumber
\end{eqnarray}
where, $\mathbf{\Lambda}^{(\rho)}=\boldsymbol{\nabla}\log\overline{\rho}$
is the inverse density stratification height,
$\mathbf{\Lambda}^{(u)}=\frac{1}{2}\boldsymbol{\nabla}\log\left(\eta_{T}^{(0)}\right)$
is the same for the turbulent diffusivity, $\bm{e}=\boldsymbol{\Omega}/\left|\boldsymbol{\Omega}\right|$
is a unit vector along the axis of rotation. The turbulent pumping,
$\gamma_{ij}^{(\Lambda)}$, depends on mean density and turbulent
diffusivity stratification, and on the Coriolis number $\Omega^{*}=2\tau_{c}\Omega_{0}$
where $\tau_{c}$ is the typical convective turnover time and $\Omega_{0}$
is the global angular velocity. Following the results of P08,
$\gamma_{ij}^{(\Lambda)}$ is expressed as follows:
\begin{eqnarray}
\gamma_{ij}^{(\Lambda)} & = & 3\eta_{T}\left\{
  f_{3}^{(a)}\Lambda_{n}^{(\rho)}+f_{1}^{(a)}\left(\bm{e}\cdot\boldsymbol{\Lambda}^{(\rho)}\right)e_{n}\right\}
\varepsilon_{inj}\label{eq:pump-1}\\
&-&
3\eta_{T}f_{1}^{(a)}e_{j}\varepsilon_{inm}e_{n}\Lambda_{m}^{(\rho)}
\nonumber \\
 & - & 3\eta_{T}\left(\varepsilon-1\right)\left\{
   f_{2}^{(a)}\Lambda_{n}^{(u)}
+f_{1}^{(a)}\left(\bm{e}\cdot\boldsymbol{\Lambda}^{(u)}\right)e_{n}\right\} \varepsilon_{inj}.\nonumber
\end{eqnarray}

The effect of turbulent diffusivity,
which is anisotropic due to the Coriolis force, is given by:
\begin{equation}
\eta_{ijk}=3\eta_{T}\left\{ \left(2f_{1}^{(a)}-f_{2}^{(d)}\right)\varepsilon_{ijk}-2f_{1}^{(a)}e_{i}e_{n}\varepsilon_{njk}\right\} .\label{eq:diff}
\end{equation}
We also include the nonlinear generation effects which is induced
by the large-scale current and the global rotation that is usually
called as the $\Omega\times J$ effect  or the $\delta$ dynamo effect \citep{rad69}.
It is supported by the numerical simulations \citep{2008A&A...491..353K,2011A&A...533A.108S}.
P08 suggested that:

\begin{equation}
\eta_{ijk}^{(\delta)}=3\eta_{T}C_{\delta}f_{4}^{(d)}e_{j}\left\{ \tilde{\varphi}_{7}^{(w)}\delta_{ik}+\tilde{\varphi}_{2}^{(w)}\frac{\overline{B}_{i}\overline{B}_{k}}{\overline{B}^{2}}\right\} ,\label{eq:delta}
\end{equation}
where, $C_{\delta}$ measures the amplitude of the $\Omega\times J$
effect, $\tilde{\varphi}_{2,7}^{(w)}\left(\beta\right)$ are normalized
versions of the magnetic quenching functions $\varphi_{2,7}^{(w)}$
given in P08. They are defined as follows, $\tilde{\varphi}_{2,7}^{(w)}\left(\beta\right)=\frac{5}{3}\varphi_{2,7}^{(w)}\left(\beta\right)$.
The last term in Eq.(\ref{eq:delta}) is the nonlinear contribution
to the $\Omega\times J$-effect. Its structure is the same as for
the $\alpha$ effect because the associated electromotive force is
proportional to $\frac{3}{2}\eta_{T}C_{\delta}f_{4}^{(d)}\tilde{\varphi}_{2}^{(w)}\overline{B}_{i}\left(\mathbf{e}\cdot\boldsymbol{\nabla}\right)\log\overline{\mathbf{B}}^{2}$
(see details in P08). Thus this effect works similar to the $\alpha$
effect that is excited by the nonlinear buoyant instability of large-scale
magnetic field. The functions $f_{\{1-11\}}^{(a,d)}$in Eqs(\ref{eq:diff})
depend on the Coriolis number. They can be found in P08 (see also,
\citealp{pk11,ps11}).

The mixing-length is defined as $\ell=\alpha_{{\rm MLT}}\left|\Lambda^{(p)}\right|^{-1}$,
where $\Lambda{}^{(p)}=\boldsymbol{\nabla}\log\overline{p}\,$ is the
inverse  pressure variation height, and $\alpha_{{\rm MLT}}=2$. The turbulent
diffusivity is parameterized in the form, $\eta_{T}=C_{\eta}\eta_{T}^{(0)}$,
where $\eta_{T}^{(0)}={\displaystyle \frac{u'^{2}\tau_{c}}{3f_{ov}\left(r\right)}}$
is the characteristic mixing-length turbulent diffusivity, $\ell$
is the typical correlation length of the turbulence, $C_{\eta}$ is
a constant to control the efficiency of large-scale magnetic field
dragging by the turbulent flow. Also, we modify the mixing-length
turbulent diffusivity by factor $f_{ov}(r)=1+\exp\left(50\left(r_{ov}-r\right)\right)$,
$r_{ov}=0.725R_{\odot}$ to get the saturation of the turbulent parameters
to the bottom of the convection zone.  The latter is suggested by
the numerical simulations (see, e.g., \citealt{oss2001,oss02,2008A&A...491..353K}).
The results do not change very much if we apply $\mathbf{\Lambda}^{(u)}=C_{v}\boldsymbol{\nabla}\log\left(\eta_{T}^{(0)}\right)$
with $C_{v}\le0.5$. For the greater $C_{v}$ we get the steady non-oscillating
dynamo which is concentrated to the bottom of the convection zone.
The purpose to introduce the additional parameters like $C_{v}=0.5$
and $f_{ov}(r)$ is to get the distribution of the $\alpha$ effect
closer to the result obtained in the numerical simulations.

 The bottom of the integration
domain is $r_{b}=0.715R_{\odot}$ and the top of the integration domain
is $r_{e}=0.99R_{\odot}$. The choice of parameters in the dynamo
is justified by our previous studies \citep{pk11mf}, where it has been
shown that solar-types dynamos can be obtained for $C_{\alpha}/C_{\delta}\ge2$.
In those papers we find an approximate threshold $C_{\alpha}\approx0.03$
for a given value of diffusivity dilution factor $C_{\eta}=0.05$. The latter
was chosen to tune the solar cycle period.
\end{document}